\documentclass[prl,twocolumn,superscriptaddress,amsmath,amssymb]{revtex4}
\usepackage{graphicx}
\usepackage{units}
\usepackage{color}
\usepackage[dvipsnames]{xcolor}
\usepackage{eqnarray}
\usepackage{tabularx}
\usepackage{bbm}
\usepackage{changes}

\def\cm{cm$^{-1}$}
\def\be{\begin{equation}}
\def\ee{\end{equation}}
\def\ba{\begin{eqnarray}}
\def\ea{\end{eqnarray}}

\begin{document}

    \title{Two-channel conduction in YbPtBi}
    \author{M. B. Schilling}
    \author{A. L\"ohle}
    \author{D. Neubauer}
    \affiliation{1.~Physikalisches Institut, Universit\"at Stuttgart, Pfaffenwaldring 57, 70569 Stuttgart, Germany}
    \author{C. Shekhar}
    \author{C. Felser}
    \affiliation{Max-Planck-Institut f\"ur Chemische Physik fester Stoffe, 01187 Dresden, Germany}
    \author{M. Dressel}
    \author{A. V. Pronin}
    \affiliation{1.~Physikalisches Institut, Universit\"at Stuttgart, Pfaffenwaldring 57, 70569 Stuttgart, Germany}
    \date{\today}

\begin{abstract}
We investigated transport, magnetotransport, and broadband optical
properties of the half-Heusler compound YbPtBi. Hall measurements
evidence two types of charge carriers: highly mobile electrons with
a temperature-dependent concentration and low-mobile holes; their
concentration stays almost constant within the investigated
temperature range from 2.5 to 300~K. The optical spectra (10~meV --
2.7~eV) can be naturally decomposed into contributions from intra-
and interband absorption processes, the former manifesting
themselves as two Drude bands with very different scattering rates,
corresponding to the charges with different mobilities. These
results of the optical measurements allow us to separate the
contributions from electrons and holes to the total conductivity and
to implement a two-channel-conduction model for description of the
magnetotransport data. In this approach, the electron and hole
mobilities are found to be around 50000 and 10~cm$^{2}$/Vs at the
lowest temperatures (2.5 K), respectively.
\end{abstract}

\maketitle

\textit{\textbf{Introduction.}} For 25 years YbPtBi was renowned as
a heavy-fermion compound that exhibits one of the highest effective
electron masses among the strongly correlated electron systems
\cite{Fisk1991, Canfield1991}. Its Kondo temperature is around 1~K;
in addition an antiferromagnetic transition is observed at 0.4~K. So
far, most of the experimental studies on YbPtBi explore its
heavy-fermion state and a possible quantum critical point; hence
they focus on temperatures below 2~K \cite{Fisk1991, Canfield1991,
Movshovich1994, Oppeneer1997, Mun2013, Ueland2014, Mun2015}.

More recently, however, it has been emphasized that YbPtBi belongs
to the large family of intermetallic ternary, so-called
half-Heusler, compounds, which demonstrate a variety of rather
interesting electronic properties. Several half-Heuslers are
predicted to exhibit band inversion at the $\Gamma$ point, leading
to topologically non-trivial states \cite{Chadov2010, Lin2010,
AlSawai2010, Shekhar2016}. Due to the combination of diverse
electronic properties and non-trivial band topology, half-Heuslers
are currently recognized as extremely promising objects in the
research towards functioning materials. This calls for a more
comprehensive look on the electronic properties of YbPtBi, beyond
the heavy-fermion state.

In our present study, we concentrate on the temperature range well
above the Kondo temperature, i.e.\ 2.5 to 300~K, where from our
magnetotransport and optical measurements we can draw conclusions on
the carrier dynamics. We present evidence that two types of charge
carriers coexist in this compound: highly mobile electrons with a
temperature-dependent carrier concentration of the order of
$10^{18}~{\rm cm}^{-3}$ and holes with a rather high and basically
temperature-independent concentration of $10^{20}~{\rm cm}^{-3}$
that possess a very low mobility. We find the values of electron
mobility in YbPtBi to be record high for half-Heuslers. The presence
of the highly mobile carriers is typical for materials with linear
bands \cite{Shekhar2015, Neupane2014}. Possible presence of such
bands in YbPtBi has been noticed e.g. in Ref.~\cite{Chadov2010}, but
still remains an open issue. The found very high mobility in YbPtBi
indicates that Dirac physics might indeed be relevant for this
compound, however more studies in this regard are certainly
necessary. \\

\textit{\textbf{Experiment.}} YbPtBi single crystals were grown by
the solution growth method, where Bi acts as a flux. Stoichiometric
quantities of freshly polished pieces of elements Yb, Pt, and Bi of
purity $>99.99$\% in the atomic ratio of 0.7:0.7:10 were put in a
tantalum crucible and then sealed in a dry quartz ampoule under
$3-5$ mbar argon pressure. The filled ampoules were heated at a rate
of 100 K/h up to 1473 K, followed by 12 hours of soaking; after that
the furnace temperature was decreased to 1373 K. For crystal growth,
the temperature was slowly reduced from 1373 K to 873 K by 2 K/h and
the surplus of Bi flux was removed by decanting the ampoule at 873
K. Using this method, we obtained $3-5$ mm regular triangular shaped
crystals, with a preferred growth in the (111)-direction. The
general crystal growth procedure was followed from
literature~\cite{Canfield1992}. The space group and lattice
parameters are found to be $F\overline{4}3m$ (cubic face-centered)
and 6.591 \AA, respectively, consistent with previous
reports~\cite{Nowotny1941, Robinson1994, Ueland2015}.

Specimens of appropriate shapes were cut from a single crystal,
e.g.\ Hall bars for (magneto)transport experiments and a
large-surface sample for optical reflectivity. Direct-current
resistivity measurements were performed in a custom-made setup by
cooling from room temperature down to 2.5~K. Transversal
magnetoresistance (MR) and Hall resistivity measurements were
conducted at the same temperatures in magnetic fields $B$ of up to
6.5~T. The voltages/currents were measured/applied within the (111)
plane and magnetic field was along the [111] axis.

The optical reflectivity $R(\nu)$ was measured from the (111) plane
in the temperature range between $T = 12$ and 300~K using two
Fourier-transform infrared spectrometers, a Bruker IFS 113v and a
Bruker Vertex 80v equipped with an infrared microscope. This way we
covered the frequency range from $\nu = \omega/(2\pi) = 100$ to
22\,000~\cm\ ($\hbar\omega = 12 - 2700$~{meV}). The sample for the
optical experiments had lateral dimensions of roughly 2 by 2~mm and
a typical thickness of 0.5~mm; all optical experiments were
performed on freshly cleaved (111) surfaces. For low frequencies
($50 - 1000$~\cm), an in-situ gold evaporation
technique~\cite{Homes1993} was utilized for reference measurements.
Freshly evaporated gold and protected-silver mirrors served as
references at higher frequencies. In accord with the cubic
face-centered crystal structure, measurements with linearly
polarized light revealed isotropic optical properties.

\begin{figure}[t]
    \centering
    \includegraphics[width=\columnwidth]{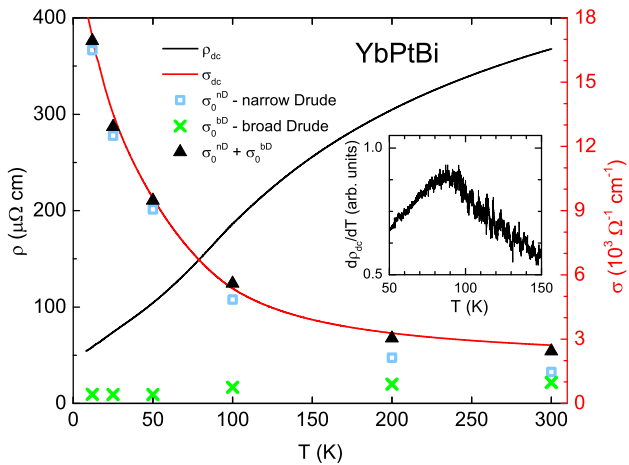}
\caption{Temperature-dependent four-point resistivity $\rho_{\rm
dc}$ (black line, left scale) and dc conductivity, $\sigma_{\rm
dc}=1/\rho_{\rm dc}$, (red line, right axis) of YbPtBi. In addition,
the bold dots correspond to the values obtained as $\nu \rightarrow
0$ extrapolations of the two Drude contributions, broad and narrow,
to the optical conductivity, and to the sum of the two, see text.
The inset shows ${\rm d}\rho_{\rm{dc}}/{\rm d}T$.}
    \label{dc}
\end{figure}

From the measured reflectivity $R(\nu)$, the optical conductivity,
$\sigma(\nu) = \sigma_{1}(\nu) + {\rm i}\sigma_{2}(\nu)$, and
dielectric function, $\varepsilon(\nu) = 1-2\rm{i}\sigma/\nu =
\varepsilon_{1}(\nu) + {\rm i}\varepsilon_{2}(\nu)$, were extracted
via the Kramers-Kronig relations \cite{DresselGruner02}. In the
course of the paper, we express our optical results in terms of
$\sigma_{1}(\nu)$ and $\varepsilon_{1}(\nu)$. For a Kramers-Kronig
analysis, the measured data have to be extrapolated to zero and high
frequencies. It turns out that in the case of our measurements the
commonly applied Hagen-Rubens extrapolation to zero frequency is not
adequate because a very narrow Drude component is present in the
spectra with a scattering rate comparable to our lowest measurement
frequency, $\nu_{\rm min} \approx 100$~\cm. Thus, we used a set of
Lorentzians instead --~some at a finite, some at zero frequency~--
to fit the measured reflectivity. Between $\nu = 0$ and $\nu_{\rm
min}$ these Drude-Lorentz fits were utilized as zero-frequency
extrapolations. On the high-frequency side ($\nu \rightarrow
\infty$), the x-ray atomic scattering functions were employed
according to Tanner \cite{Tanner2015}. \\

\textit{\textbf{Magnetotransport.}} The dc resistivity $\rho_{\rm
dc}$ and conductivity, $\sigma_{\rm dc}=1/\rho_{\rm dc}$, are
displayed in Fig.~\ref{dc} versus temperature. Upon cooling,
$\rho_{\rm dc}(T)$ continuously decreases; the residual resistivity
ratio is 8, comparable to the values reported earlier
\cite{Fisk1991, Mun2013}. In $\rho_{\rm dc}(T)$, a point of
inflection, tentatively attributed to the influence of
crystalline-electric-field effects \cite{Mun2013}, is observed
around $T=(90 \pm 5)$~K.

\begin{figure}[b]
    \centering
    \includegraphics[width=\columnwidth]{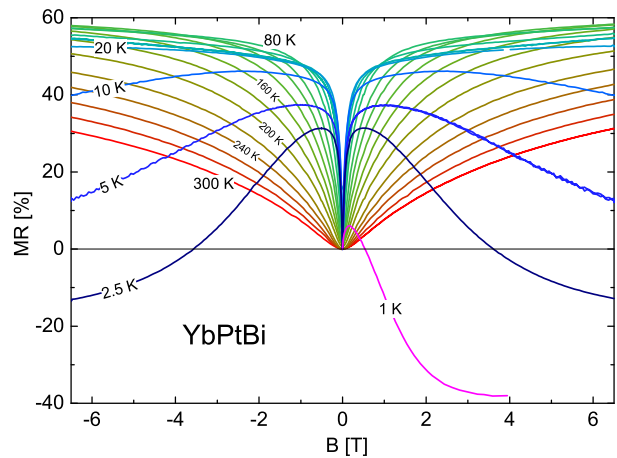}
\caption{Magnetoresistance of YbPtBi as measured between 2.5 and
300~K. The 1~K data are extracted from Ref.~\onlinecite{Mun2013}.}
    \label{MR}
\end{figure}

Fig.~\ref{MR} shows the results of our magnetotransport experiments
by plotting the field-dependent resistivity taken at different
temperatures. The MR curves are supplemented by data of Mun {\it et
al.} \cite{Mun2013} taken at $T=1$~K. At all temperatures the
magnetoresistance sharply increases with $B$ in the range of low
magnetic fields. It then flattens out for elevated temperatures,
eventually reaching almost 60\%\ as $B \rightarrow 6.5$~T. Such
sharp increase in MR at low fields and its high-field saturation can
be interpreted as localization of carriers with low scattering rate
(i.e. with high mobility) in the fields of $\sim 0.5$ T, while other
type of carriers, with high scattering rates and low mobility, still
provides the dc transport. For $T < 20$~K the magnetoresistance is
not monotonic anymore: a pronounced maximum is observed that shifts
to lower fields as temperature decreases. At very low temperatures,
$T\leq 2.5$ K, MR changes sign: at $B=3.5$~T for $T = 2.5$~K and
around 0.5~T for the 1~K data of Mun {\it et al.} \cite{Mun2013}.
The position of the maximum in the MR-versus-$B$ curves follows a
power-law temperature dependence, $T^{n}$, with $n = 1.2 \pm 0.1$.
Negative MR in YbPtBi is usually discussed in terms related to Kondo
physics \cite{Fisk1991, Mun2013}, though no in-depth discission is
available so far. It is worth to note here, that similar, but
somewhat different, results on transversal MR are reported recently
for related compounds, ScPtBi \cite{Hou2015} and HoPdBi
\cite{Pavlosiuk2016}. In ScPtBi, no negative MR is observed in the
whole range of measured fields ($0 - 10$ T) and temperatures ($2 -
300$ K), but the shape of the MR-versus-$B$ curves is very similar
to our higher-temperature results. In HoPdBi, the negative MR sets
in already at 50 K in 7 T and the overall shape of the lower
temperature MR curves are similar to our results for $T < 20$ K, but
the temperature dependencies of the MR maximum is different.

\begin{figure}
    \centering
    \includegraphics[width=\columnwidth]{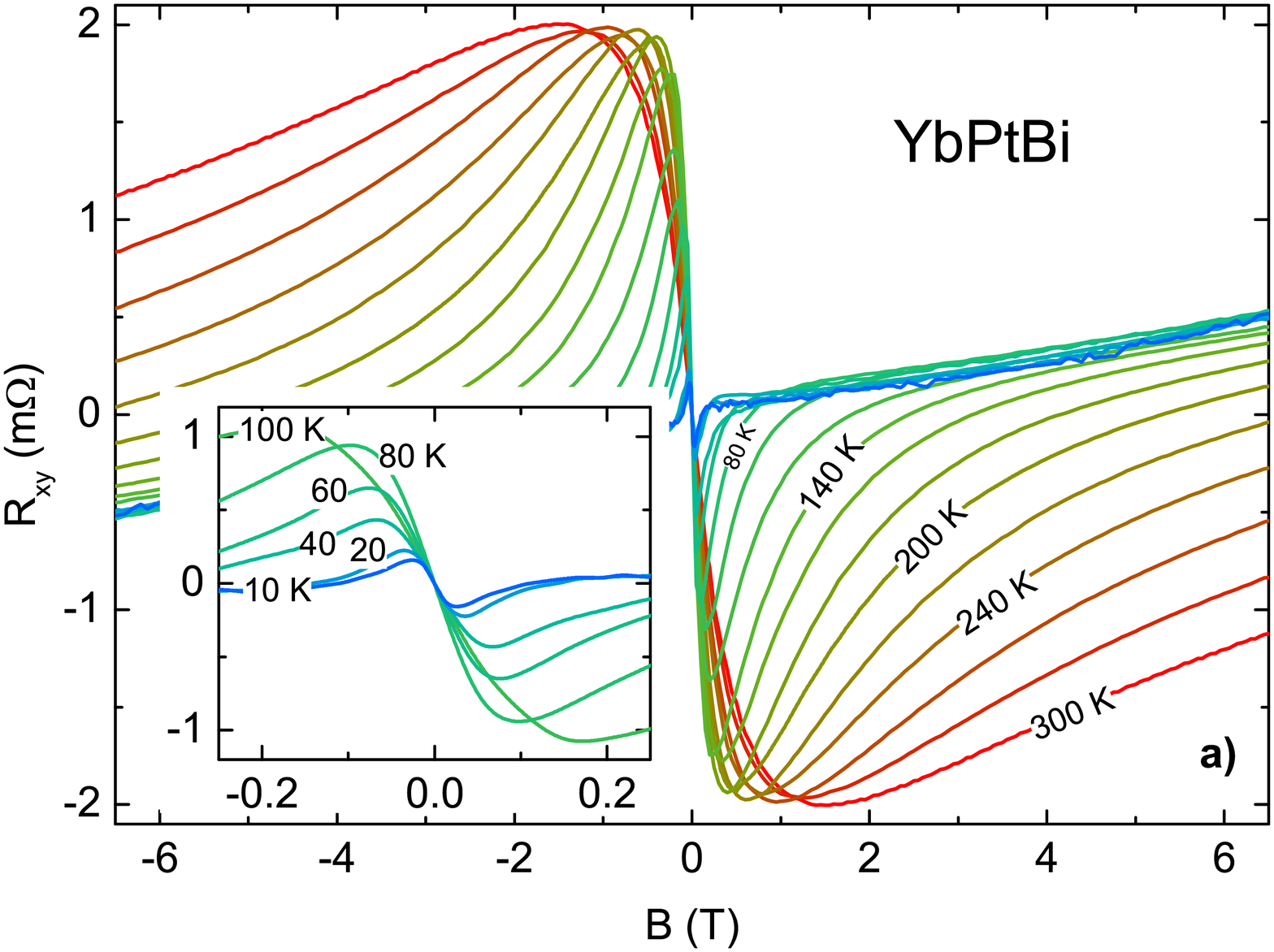}
    \includegraphics[width=\columnwidth]{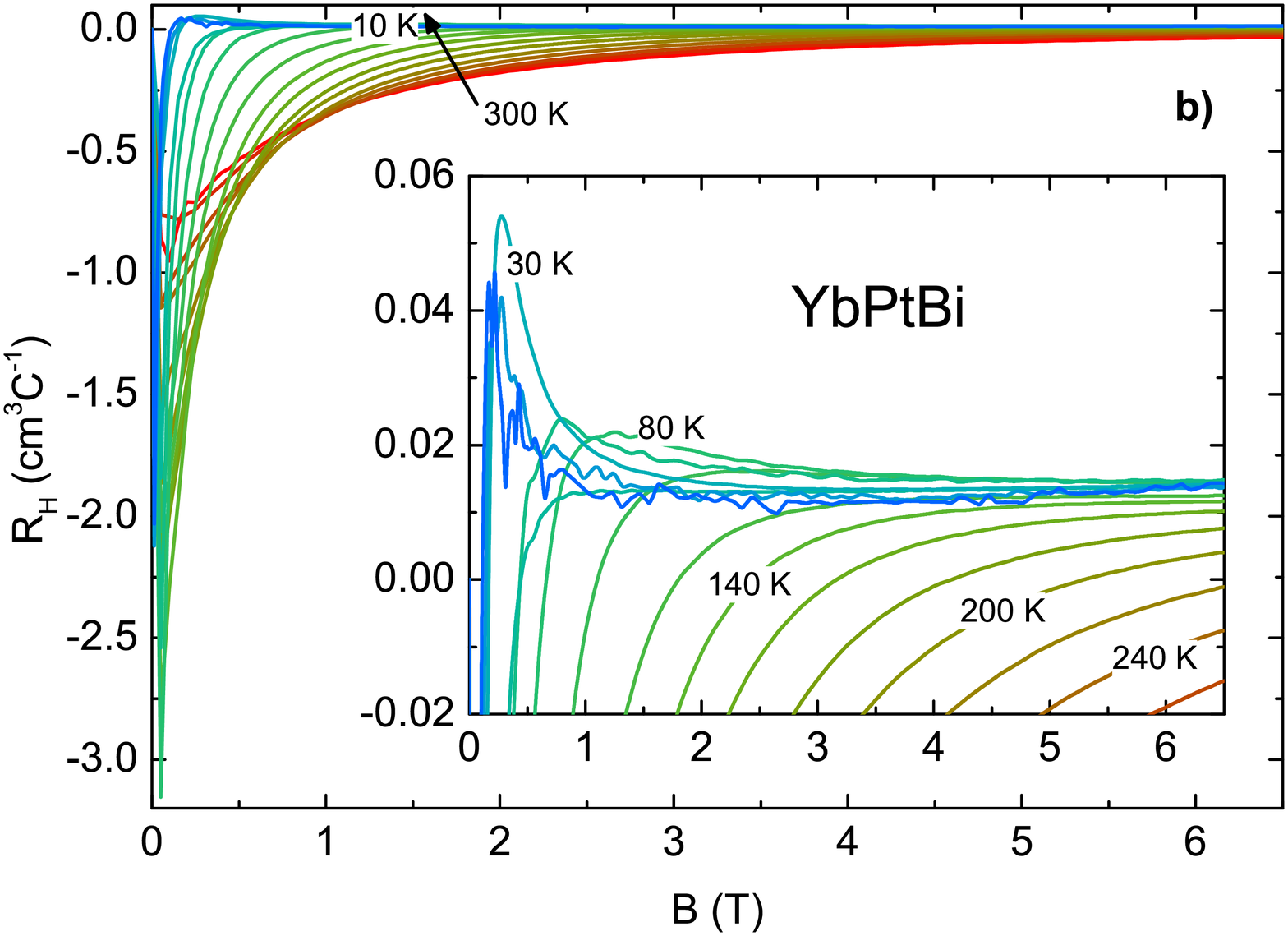}
\caption{(a)~Hall resistance and (b) Hall coefficient (bottom) of
YbPtBi at 2.5 to 300~K. The insets zoom-in the Hall resistance at
low fields (top panel) and the small values of the Hall coefficient
(bottom panel).}
    \label{Hall}
\end{figure}

The results of the Hall measurements on YbPtBi, i.e.\ the Hall
resistance, $R_{xy}(B,T)$, and the Hall coefficient, $R_H(B,T) =
\rho_{xy}/B$, are plotted in Fig.~\ref{Hall} as the functions of
applied magnetic field for various temperatures as indicated. The
low-field behavior is magnified in the inset in order to demonstrate
that $R_H$ is always negative in low fields. As the field increases,
the Hall coefficient eventually changes sign: electrons get
localized and the Hall response becomes dominated by holes
(accordingly, the MR saturates, as discussed). Within our range
($B<6.5$~T), $R_H$ turns positive for all temperatures below 260~K
and saturates at $0.015~{\rm cm}^{3}{\rm C}^{-1}$. Obviously, this
very value would be reached at all temperatures up to 300~K, if we
could go to higher magnetic fields, as demonstrated in the inset of
the bottom panel. Thus, we conclude that the hole concentration
$n_{h}$ is temperature independent and equal to $(5.2 \pm 0.6)
\times10^{20}~{\rm cm}^{-3}$.

\begin{figure}
    \centering
\includegraphics[width=\columnwidth]{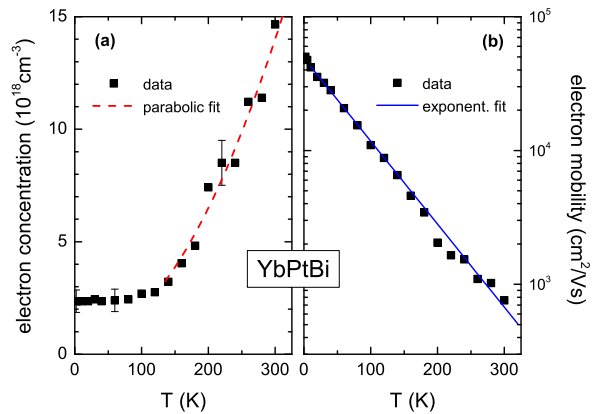}
\caption{Electron concentration (a) and mobility (b) in YbPtBi as
obtained from Hall measurements with the correction from the
optical-conductivity fits, as described in the text. The hole
concentration and mobility are equal to $5.2\times10^{20}~{\rm
cm}^{-3}$ and $\sim 10~{\rm cm}^{2}/{\rm Vs}$, respectively, and
both are basically temperature independent. At higher temperatures,
the electron concentration follows a $T^{2}$ behavior, as indicated
by the dashed line, while the election mobility can nicely be fitted
in the entire temperature range by an exponent shown as the solid
line.}
    \label{concentration}
\end{figure}

Oppositely, the electron concentration $n_{e}$ demonstrates a strong
temperature dependence. We calculate $n_{e}(T)$ from $R_H = 1/(n_{e}
\cdot e)$, taking into account only the initial slop in
$\rho_{xy}(B)$ at low fields, and plot the results in
Fig.~\ref{concentration}(a).

The two-carrier-type scenario for YbPtBi does not allow us to
calculate the carrier mobility in a straightforward fashion because
at this stage we cannot separate the contributions of each carrier
type to the total conductivity $\sigma_{\rm dc}$. In the so-called
one-carrier-type (OCT) approach (i.e. if we use the total measured
$\sigma_{\rm dc}$ to compute both, electron and hole, mobilities
from the corresponding Hall coefficients), one can though estimate
roughly the mobilities: $\mu^{\rm OCT}_{e}=R_{H}^{e}\cdot\sigma_{\rm
dc}\approx53\,000~{\rm cm}^{2}/{\rm Vs}$ and $\mu^{\rm
OCT}_{h}=R_{H}^{h}\cdot\sigma_{\rm dc}\approx200~{\rm cm}^{2}/{\rm
Vs}$ at $T=10$~K. This estimate not only provides the upper limits
for the mobilities, but also shows that the mobilities of electrons
and holes in YbPtBi differ from each other by orders of magnitude.
Below we will describe, how our optical measurements provide
additional information in order to separate the electron and hole
contributions to the conductivity and thus to get more accurate
values of the carrier mobilities. \\

\begin{figure}
    \centering
    \includegraphics[width=\columnwidth]{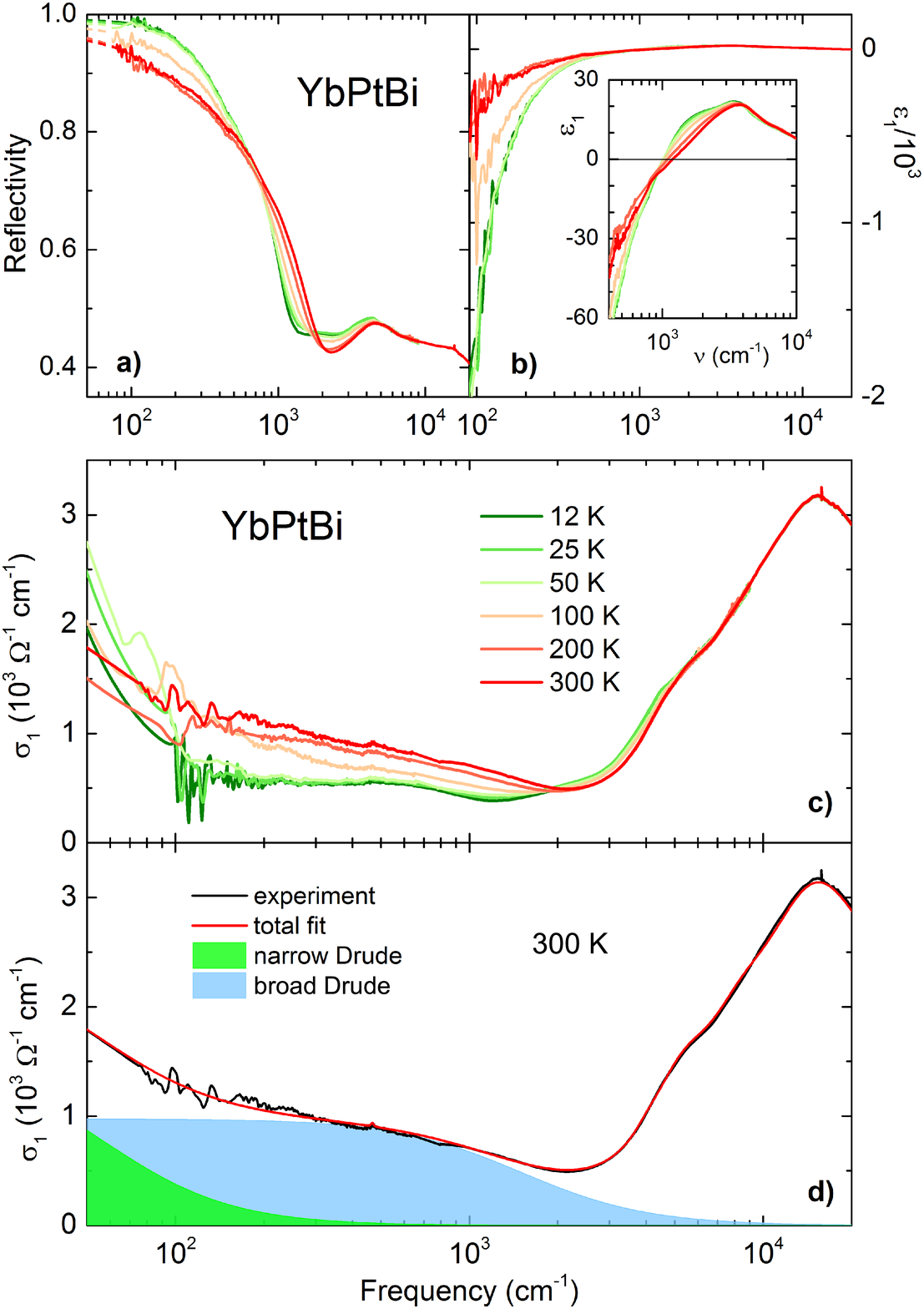}
\caption{Frequency-dependent reflectivity (a), dielectric constant
(b), and optical conductivity (c) of YbPtBi for selected
temperatures between 12 and 300~K. The inset in panel (b) zooms in
$\varepsilon_{1}(\nu)$ near the zero crossing. Panel (d): fit of the
optical conductivity at 300~K using a Drude-Lorentz model. The
intraband conductivity can be represented with two Drude components.
The reflectivity and dielectric constant are fitted using the same
Drude-Lorentz model (fits are not shown).}
    \label{optics}
\end{figure}

\textit{\textbf{Optics.}} Fig.~\ref{optics} displays the results of
our optical investigations. The measured reflectivity spectra
$R(\nu)$ are plotted in panel (a) for various temperatures. The
Drude-like interband contributions below 2000~\cm\ are separated
from the interband transitions at higher frequencies. The
characteristic downturn in $R(\nu)$ near the plasma edge ($\sim$
1000~\cm), however, reaches only 0.45, indicating some overlap of
the intra- and interband contributions in the mid-infrared region.

Panel (c) of Fig.~\ref{optics} shows the optical conductivity as
derived from the Kramers-Kronig analysis of the reflectivity data.
Here not only the absorption bands due to the intra- and interband
transitions can be separated; it also becomes apparent that the
low-frequency response consists of two Drude-like terms.

The dielectric constant $\varepsilon_1(\nu)$ is presented in
Fig.~\ref{optics}(b). It is negative below approximately 1000~\cm\,
due to the free-carrier (Drude) contributions and reveals a broad
maximum at around 3500~\cm\ signaling an interband absorption edge
at somewhat lower frequencies \cite{Yu2010, DresselGruner02} and
confirming the assignment of the optical-conductivity modes above
2000~\cm\ to interband transitions. At $T<100$~K, i.e. in the same
temperature range, where $\rho_{\rm dc}(T)$ demonstrates the
inflection point, a shoulder in $\varepsilon_1(\nu)$ appears at
approximately 1600~cm$^{-1}$. The shoulder becomes more pronounced
as the temperature decreases. It likely corresponds to an additional
interband transition and hence indicates a possible band-structure
modification at around $T \approx 90$~K \cite{optics}.

Zero-line crossings of the $\varepsilon_1(\nu)$ curves (the inset of
Fig.~\ref{optics}(b)) shift towards higher frequencies as $T$
increases. This reflects the increased carrier concentration at
higher temperatures: the zero-line crossings of $\varepsilon_1(\nu)$
can be taken as a measure of the screened plasma frequency,
$\omega_{pl}^{scr}/2\pi$, which in turn is proportional to the
carrier concentration.

In order to get more insight into the charge carrier dynamics, we
simultaneously fit the experimental spectra of $\sigma_1(\nu)$,
$\varepsilon_1(\nu)$, and $R(\nu)$ for each temperature by the
Lorentz-Drude model. Results of these fits are exemplified in the
panel (d) for $\sigma_{1}$ at $T=300$ K. For the entire temperature
range of our measurements, we were not able to describe the observed
conductivity with a single Drude term, but obtained a satisfactory
description only for two such contributions. These Drude terms have
very different (by more than an order of magnitude) scattering
rates, $1/\tau$. Analysis of Shubnikov--de~Haas oscillations in
YbPtBi yields the effective carrier masses $m^{*}$ close to the
free-electron mass $m_{e}$ for any band, ranging from $\sim 0.5$ to
$1.5m_{e}$ \cite{Mun2015}. Thus, the large difference between the
electron and hole mobilities in YbPtBi should be mostly related to
the difference in the scattering rates ($\mu=e\tau/m^{*}$) and one
can ascribe the Drude term with the smaller scattering rate (the
narrow Drude) to the highly mobile electrons and the term with
larger scattering rate (the broad Drude) to the holes with low
mobility.

The optical fits with the two Drude terms allow us to separate the
contributions from the electrons and holes into the total
conductivity. The bold dots in Fig.~\ref{dc} correspond to the
values obtained as $\nu \rightarrow 0$ extrapolations of the two
Drude contributions. As one can see from the Figure, their sum
nicely follows the dc-conductivity curve. It is also apparent that
at low temperatures electrons dominate the conductivity, while as $T
\rightarrow 300$ K, the contributions of elections and holes to
$\sigma_{\rm dc}$ become comparable, the electrons still providing a
larger contribution.

Using the interpolations between the points, obtained from the
optical-data fits, and the carrier concentrations, obtained from the
Hall measurements, we calculate the carrier mobilities of electrons,
$\mu_{e}$, and holes, $\mu_{h}$. The electron mobility is plotted in
Fig.~\ref{concentration}(b). As the hole contribution to the dc
conductivity is (much) lower than the electron contribution
(Fig.~\ref{dc}), the values for electron motility do not differ much
from the values, obtained using the one-carrier-type approximation.
However, the hole mobility, calculated with the use of the optical
fits, is much lower than the one obtained in the one-carrier-type
approach. We find $\mu_{h}=(10\pm5)~{\rm cm}^{2}/{\rm Vs}$, being
basically temperature independent. However, we cannot exclude a weak
temperature dependence of $\mu_{h}$ within the given error bar: the
hole scattering rate changes somewhat as a function of $T$.

As one can see from Fig.~\ref{concentration}(b), the electron
mobility shoots up exponentially as $T$ decreases \cite{oct}. We
could fit $\mu_{e}$ with $\mu_{e}(T)=\mu_{e}(0){\rm exp}(-T/T_{0})$
in the entire temperature range. We have found
$\mu_{e}(0)=50\,000~{\rm cm}^{2}/{\rm Vs}$ and $T_{0}=70$ K. The
exponential behavior of $\mu_{e}$ is quite remarkable: usually, even
in the materials with the highest reported mobilities, such as e.g.
the Weyl semimetal NbP, $\mu(T)$ saturates at temperatures below
some $20 - 30$ K \cite{Shekhar2015}. The exponential behavior of
$\mu_{e}(T)$ in YbPtBi persists down to our lowest temperature (2.5
K) despite leveling off the electron concentration at $T < 100$~K,
Fig.~\ref{concentration}(a). This signals collapsing of the electron
scattering in YbPtBi at low temperatures. The electron concentration
as a function of temperature does not show a sign of Arrhenius
behavior at any $T$, the flat behavior at low temperatures being
followed by a power law, $\mu_{e} \propto T^{2}$, at $T > 100$~K. \\


\textit{\textbf{Conclusions.}} From our transport, magnetotransport,
and optical investigations of the half-Heusler compound YbPtBi at $T
\geq 2.5$ K (i.e. not in the heavy-fermion state), we can separate
two conduction channels caused by electrons and holes. Electrons
posses a temperature-dependent concentration, high mobility, and low
scattering rates. Holes form the second channel with the
concentration and mobility that are basically temperature
independent. While at high temperatures, i.e.\ $200 - 300$~K, both
channels provide comparable contributions to the dc transport, at
low temperatures the electron channel dominates overwhelmingly. This
is due to very high mobility, $\mu_{e}(T = 2.5 \rm{K)}=50\,000~{\rm
cm}^{2}/{\rm Vs}$, and low scattering of the electrons at $T <
100$~K. To the best of our knowledge, the values of electron
mobility in YbPtBi, are record high for half-Heusler compounds. \\


\textit{\textbf{Acknowledgements.}} We thank Hongbin Zhang and
Binghai Yan for useful discussions. Gabriele Untereiner and Roland
R\"osslhuber provided valuable experimental support. This work was
funded by the Deutsche Forschungsgesellschaft (DFG) via DR228/54-1.


\begin{thebibliography}{99}
\bibitem{optics} More details on the interband portion of the optical conductivity,
    as well as a comparison of our optical results with the optical
    spectra, computed from band-structure calculations, will be given
    elsewhere.
\bibitem{oct} Let us note, that $\mu_{e}^{\rm OCT}$ demonstrates the
    same exponential behavior as $\mu_{e}$, with $\mu_{e}(0)$ and $T_{0}$ being
    only slightly changed.
\end{thebibliography}
\end{document}